\documentclass{aa}
\usepackage{graphicx}
\usepackage{txfonts}
\begin{document}

   \title{Transverse oscillation of a coronal loop induced by a flare-related jet}

   \author{J. Dai\inst{1,2}
          \and
          Q. M. Zhang\inst{1,2,3}
          \and
          Y. N. Su\inst{1,2}
          \and
          H. S. Ji\inst{1,2}
          }

   \institute{Key Laboratory of Dark Matter and Space Astronomy, Purple Mountain Observatory, CAS, Nanjing 210023, PR China \\
         \email{zhangqm@pmo.ac.cn}
         \and
         School of Astronomy and Space Science, University of Science and Technology of China, Hefei 230026, PR China \\
         \and
         State Key Laboratory of Lunar and Planetary Sciences, Macau University of Science and Technology, Macau, PR China \\
          }

   \date{Received; accepted}
    \titlerunning{Transverse oscillation of a coronal loop}
    \authorrunning{Dai et al.}

\abstract
{Kink oscillations in coronal loops are ubiquitous, and the observed parameters of oscillations are applied to estimate the magnetic field strength of the loops.}
{In this work, we report our multiwavelength observations of the transverse oscillation of a large-scale coronal loop with a length of $\geq$350 Mm.
The oscillation was induced by a blowout coronal jet, which was related to a C4.2 circular-ribbon flare (CRF) in AR 12434 on 2015 October 16.
We aim to determine the physical parameters in the coronal loop, including the Alfv\'{e}n speed and magnetic field strength.}
{The jet-induced kink oscillation was observed in extreme-ultraviolet (EUV) wavelengths by the Atmospheric Imaging Assembly (AIA) on board the Solar Dynamics Observatory (SDO).
Line of sight magnetograms were observed by the Helioseismic and Magnetic Imager (HMI) on board SDO.
We took several slices along the loop to assemble time-distance diagrams, and used an exponentially decaying sine function to fit the decaying oscillation.
The initial amplitude, period, and damping time of kink oscillation were obtained.
Coronal seismology of the kink mode was applied to estimate the Alfv\'{e}n speed and magnetic field strength in the oscillating loop.
In addition, we measured the magnetic field of the loop through non-linear force-free field (NLFFF) modeling using the flux rope insertion method.}
{The oscillation is most pronounced in AIA 171 and 131 {\AA}.
The oscillation is almost in phase along the loop with a peak initial amplitude of $\sim$13.6 Mm, meaning that the oscillation belong to the fast standing kink mode.
The oscillation lasts for $\sim$3.5 cycles with an average period of $\sim$462 s and average damping time of $\sim$976 s. The values of $\tau/P$ lie in the range of 1.5$-$2.5.
Based on coronal seismology, the Alfv\'{e}n speed in the oscillating loop is estimated to be $\sim$1210 km s$^{-1}$.
Two independent methods are applied to calculate the magnetic field strength of the loop,
resulting in 30$-$43 G using the coronal seismology and 21$-$23 G using the NLFFF modeling, respectively.}
{The magnetic field strength estimated using two different approaches are in the same order of magnitude,
which confirms the reliability of coronal seismology by comparing with the NLFFF modeling.}

\maketitle

\section{Introduction} \label{s-intro}
Waves and oscillations are prevalent in the fully ionized solar corona with temperatures of several million Kelvin (MK) \citep[see][and refererence therein]{naka05}.
The periodic transverse displacements of coronal loops are usually considered as kink oscillations detected in the extreme ultraviolet (EUV) wavelengths \citep{and09,rud09}.
Standing fast kink-mode oscillations were initially discovered by the Transition Region and Coronal Explorer \citep[TRACE;][]{hand99} mission in 171 {\AA} \citep{asch99,naka99,sch02}.
Since the launch of the Solar Dynamics Observatory \citep[SDO;][]{pesn12}, kink oscillations of coronal loops observed by
the Atmospheric Imaging Assembly \citep[AIA;][]{lemen12} on board SDO have been extensively investigated \citep[e.g.,][]{as11,wht12,ver13a,ver13b,pas16,nis17,duck18,duck19,nev19}.

The commencement of the loop oscillation usually coincides with a nearby eruption (flare, jet, or filament eruption) in the lower corona,
which is considered as the predominant mechanism for exciting kink oscillations \citep{zim15}.
After the excitation, kink oscillations experience attenuation and last for several cycles in most cases \citep{god16a,god16b}.
Resonant absorption, as a result of resonance within a finite thin layer, is believed to play a key role in the rapid damping of fast-mode kink oscillations \citep{goo02,rud02}.
Phase mixing with anomalously high viscosity is also important in the dissipation of energy during loop oscillations \citep{ofm02}.
Small-amplitude, transverse oscillations of coronal loops without significant damping have been noticed \citep{anf13,nis13,lid18,zhang20}.
The observed loop oscillations in combination with magnetohydrodynamics (MHD) wave theory provide an effective tool to determine the local physical parameters,
such as the magnetic field and Alfv\'{e}n speed of the oscillating loops, which are difficult to measure directly \citep{ed83,naka01,asch02,ver06,arr07,goo08,van08,ant11,yuan16,li17}.

Circular-ribbon flares (CRFs) are a special type of flares, whose short, inner ribbons are surrounded by circular or elliptical ribbons \citep{mas09,chen19,zqm16b,zqm19,lee20,liu20}.
Transverse loop oscillations excited by CRFs with periods of $\la$4 min have been observed by AIA \citep{zqm15,lit18}.
Recently, \citet{zqm20} investigated the transverse oscillations of an EUV loop excited by two successive CRFs on 2014 March 5. The oscillations are divided into two stages in their development:
the first-stage oscillation triggered by the C2.8 flare is decayless with lower amplitudes, and the second-stage oscillation triggered by the M1.0 flare is decaying with larger amplitudes.
The authors also estimated the magnetic field and thickness of the inhomogeneous layer of the oscillating loop with a length of $\sim$130 Mm.

In this paper, we report our multiwavelength observations of the transverse oscillation of a large-scale coronal loop excited by a blowout jet associated with the C4.2 CRF
in active region (AR) 12434 on 2015 October 16. \citet{zqm16c} studied the explosive chromospheric evaporation at the inner and outer flare ribbons using spectroscopic observations.
This work is building on the work of \citet{zqm16c} (hereafter Paper I),
and the main purpose is to estimate the magnetic field of the oscillating loop using two independent approaches, coronal seismology and magnetic field extrapolation.
This paper is organized as follows. Observations and data analysis are presented in Sect.~\ref{s-obs}.
The results are presented in Sect.~\ref{s-res}. A brief summary and discussion are presented in Sect.~\ref{s-sum}.

\section{Observations and Data Analysis} \label{s-obs}

\subsection{Instruments}
The transverse oscillation of the coronal loop was observed by SDO/AIA, which has a spatial resolution of 1$\farcs$2 and time cadence of 12 s in EUV wavelengths.
The photospheric line-of-sight (LOS) magnetograms were observed by the Helioseismic and Magnetic Imager \citep[HMI;][]{schou12} on board SDO
with a spatial resolution of 1$\farcs$2 and time cadence of 45 s.
The level\_1 data from AIA and HMI were calibrated using the standard Solar SoftWare (SSW) programs \texttt{aia\_prep.pro} and \texttt{hmi\_prep.pro}, respectively.
Soft X-ray (SXR) light curves of the flare were recorded by the GOES spacecraft with a cadence of $\sim$2.05 s.

\subsection{DEM analysis} \label{s-dem}
The differential emission measure (DEM) analysis is a useful tool to perform temperature diagnostics.
Several algorithms have been proposed and validated \citep[e.g.][]{web04,han12,asch13,plow13,mc15,su18b,mor19}.
The observed flux $\mathit{F}_i$ of each optically thin passband $i$ is determined by:
\begin{equation} \label{eqn-1}
  F_{i} = \int_{T_1}^{T_2} \mathit{R}_{i} (\mathit{T})\mathrm{DEM}(\mathit{T}) \mathit{dT}
  \,,
\end{equation}
where $\mathit{R}_i(T)$ is the temperature response function of passband $i$, and DEM($\mathit{T}$) represents the DEM of multithermal plasma as a function of temperature.
$\log T_1=5.5$ and $\log T_2=7.5$ stand for the lower and upper limits for the integral.
To carry out the inversion of DEM profile, we use the standard SSW program \texttt{xrt\_dem\_iterative2.pro} and six EUV passbands (94, 131, 171, 193, 211, and 335 {\AA}).
The method has been strictly justified and successfully applied to the temperature estimation of EUV hot channel as well as coronal jets \citep{cx12,zqm14,zqm16a}.
Note that the background emissions should be removed before inversion.

The DEM-weighted average temperature $\bar{T}$ is defined as \citep{cx12}:
\begin{equation} \label{eqn-2}
  \bar{T} = \frac{\int_{T_1}^{T_2} \mathrm{DEM}(\mathit{T}) \mathit{TdT}}{\int_{T_1}^{T_2} \mathrm{DEM}(\mathit{T})\mathit{dT}}.
\end{equation}
Then, the total column emission measure (EM) along the LOS is expressed as:
\begin{equation} \label{eqn-3}
  \mathrm{EM} = \int_{T_1}^{T_2} \mathrm{DEM}(\mathit{T})\mathit{dT}.
\end{equation}

\subsection{Flux rope insertion method} \label{s-su}
We use the flux rope insertion method developed by \citet{van04} to reconstruct the non-linear force-free field (NLFFF) of AR 12434.
The advantage of this method is that it can be applied to many different situations including both ARs \citep{su09,su11,su18a} and quiet Sun \citep{su15}
since no vector field observations are required and the magnetic field lines of the best-fit model match well the observed coronal non-potential structures.
\citet{su19} gives a detailed review on the application of the method. Reconstructing the coronal magnetic fields in the target AR requires four steps:
\begin{enumerate}
\item Extrapolating the potential field based on the corresponding photospheric LOS magnetogram.
\item According to observations, creating a cavity in the potential field model, and then inserting a magnetic flux rope along the selected paths.
\item Creating a grid of models by adjusting axial flux and poloidal flux of the inserted magnetic flux rope.
\item Starting magneto-frictional relaxation \citep{yang86} to drive the magnetic field towards a force free state, and then comparing with observations to find the best-fit model.
\end{enumerate}

\section{Results} \label{s-res}
\subsection{Transverse coronal loop oscillation}
\begin{figure*}
   \includegraphics[width=14cm,clip=]{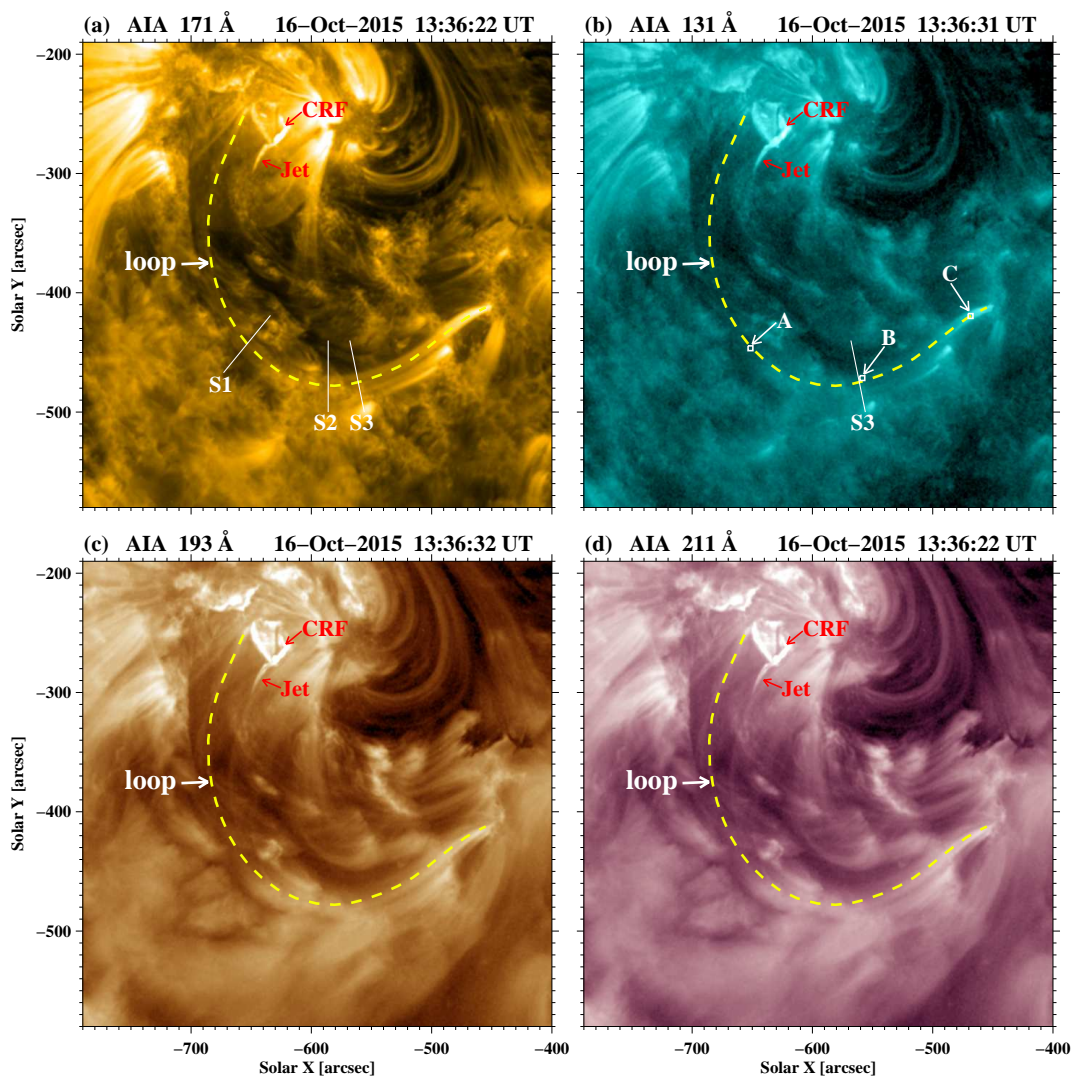}
   \centering
   \caption{EUV images of the large-scale coronal loop (yellow dashed line) observed by SDO/AIA in 171, 131, 193, and 211 {\AA} before flare.
   The straight solid lines (S1, S2, S3) are used to analyze the transverse oscillation.
   The red arrows point to the locations of circular-ribbon flare (CRF) and blowout coronal jet.
   In panel (b), the three tiny boxes along the loop mark the locations for DEM analysis.}
    \label{fig1}
\end{figure*}

Figure~\ref{fig1} shows the EUV images observed by AIA in 171, 131, 193, and 211 {\AA} before flare.
The northeast footpoint of the large-scale coronal loop (yellow dashed line) is rooted in AR 12434
and is very close to the C4.2 CRF pointed by the red arrow. As is described in Paper I, the flare brightened up from $\sim$13:36:30 UT.
The accompanying blowout jet started to rise at $\sim$13:39 UT and propagated in the southeast direction at a speed of $\sim$300 km s$^{-1}$.
Transverse oscillation of the long loop was excited by the jet and lasted for a few cycles (see the online movie \textit{oscillation.mov}).

\begin{figure}
\includegraphics[width=8cm,clip=]{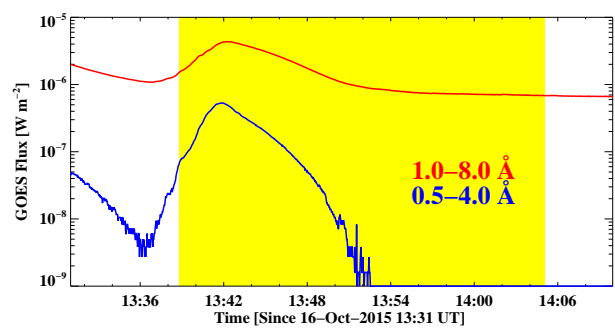}
\centering
\caption{SXR light curves of the flare in 1$-$8 {\AA} (red line) and 0.5$-$4 {\AA} (blue line).
The yellow area represents the time of loop oscillation during 13:39$-$14:05 UT.}
\label{fig2}
\end{figure}

Figure~\ref{fig2} shows the SXR light curves of the flare in 1$-$8 {\AA} (red line) and 0.5$-$4 {\AA} (blue line).
The SXR emissions started to increase at $\sim$13:36:30 UT and reached the peak values at $\sim$13:42:30 UT, before declining gradually until $\sim$13:51 UT (see also Fig. 5 in Paper I).
The time of loop oscillation during 13:39$-$14:05 UT is labeled with yellow area. It is found that the start time of transverse loop oscillation coincides with the fast ejection of jet
(see also Fig. 4 in Paper I), and the oscillation covers part of the impulsive and whole decay phase of the flare, lasting for $\sim$26 min.
The excitation of loop oscillation in our study can be interpreted by the schematic cartoon \citep[][see their Fig. 2]{zim15}.

\begin{figure}
\includegraphics[width=8cm,clip=]{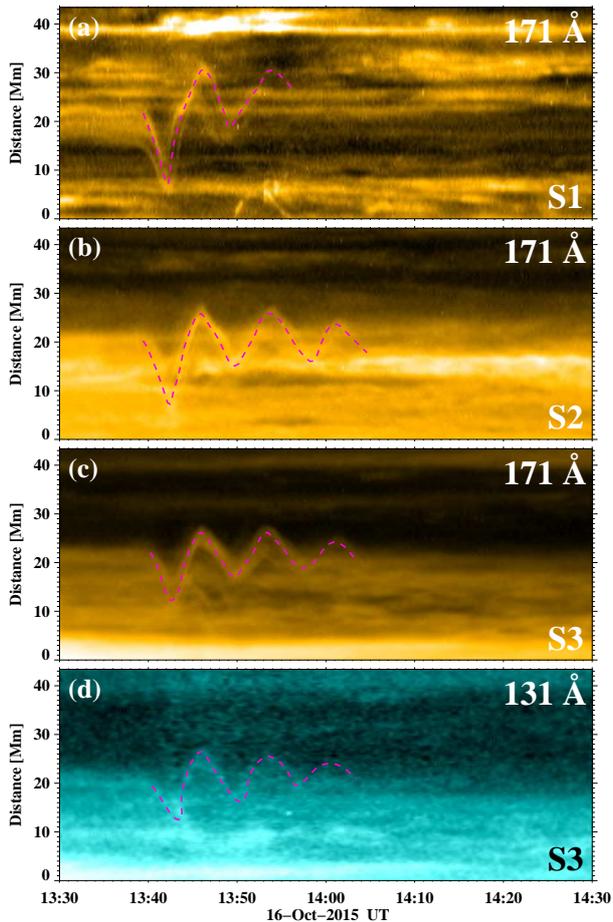}
\centering
\caption{Time-distance diagrams of the three slices (S1, S2, S3) in 171 and 131 {\AA}.
The magenta dashed lines outline the transverse loop oscillation during 13:39$-$14:05 UT.
On the $y$-axis, $s=0$ and $s=43.5$ Mm denote the southern and northern endpoints of slices, respectively.}
\label{fig3}
\end{figure}

To investigate the transverse loop oscillation, we choose three points along the loop: the first one is near the loop top, the third one is close to the southwest footpoint, and the second one in between.
We place three artificial slices (S1, S2, and S3) across the points and just perpendicular to the loop, which are drawn with white solid lines in Fig.~\ref{fig1}.
The corresponding time-distance diagrams in 171 {\AA} are plotted in the top three panels of Fig.~\ref{fig3}.
It is seen that excited by the jet, the loop deviated from the equilibrium state and began to move southward at $\sim$13:39 UT coherently.
Then the loop moved backward and oscillated with decaying amplitude for more than three cycles.
The almost identical phases of the transverse loop oscillation at different positions along the loop indicate that the oscillation belong to the standing fast kink mode.

To determine the parameters of kink oscillation along different slices, we mark the positions of the loop manually, which are connected with magenta dashed lines in Fig.~\ref{fig3}.
The kink oscillation is fitted with an exponentially decaying sine function \citep{naka99,zqm20} using the standard SSW program \texttt{mpfit.pro}:
\begin{equation} \label{eqn-4}
  A(t) = A_{0}\sin \bigg(\frac{2\pi t}{P} + \psi \bigg)e^{-\frac{t}{\tau}} + A_{1}t + A_{2},
\end{equation}
where $A_0$ is the initial amplitude, $P$ is the period, $\tau$ is the damping time, $\psi$ is the initial phase, and $A_{1}t+A_2$ is a linear term of the equilibrium position of the loop.

\begin{figure}
\includegraphics[width=8cm,clip=]{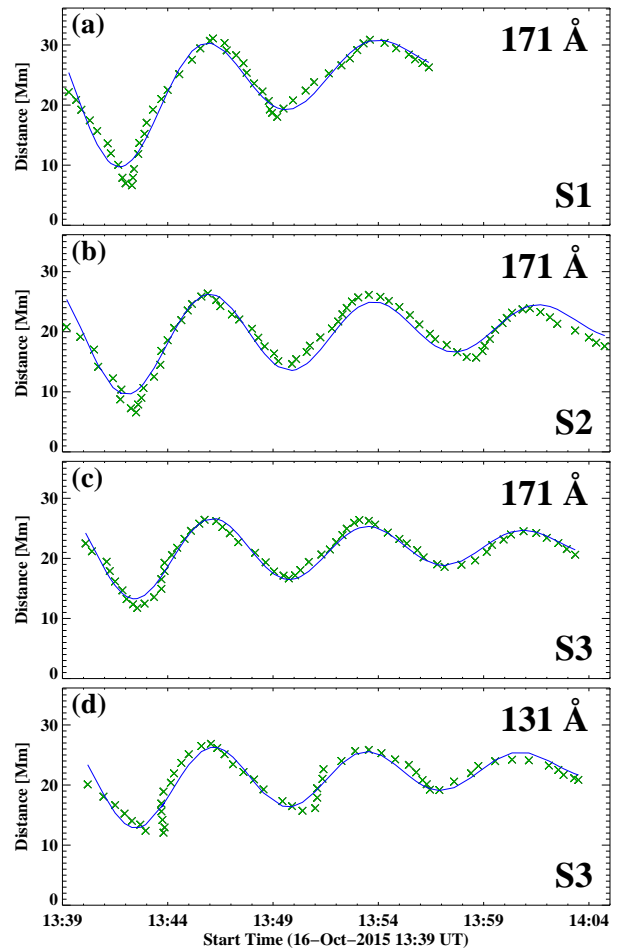}
\centering
\caption{Positions of the oscillating loop along the three slices (green crosses) and the results of curve fitting (blue lines) using Equation~\ref{eqn-4}.}
\label{fig4}
\end{figure}

In Fig.~\ref{fig4}, the green crosses represent the positions of the loop along the three slices in 171 and 131 {\AA}, and the results of curve fitting are plotted with blue lines.
It is obvious that the curve fitting using Equation~\ref{eqn-4} is satisfactory. The fitted parameters are listed in Table~\ref{tab-1}.
In 171 {\AA}, the initial amplitude decreases from $\sim$13.6 Mm near the loop top to $\sim$8.9 Mm near the footpoint.
The period ranges from $\sim$440 s to $\sim$480 s with an average value of $\sim$462 s. The damping time ranges from $\sim$710 s to $\sim$1190 s with an average value of $\sim$976 s.
The quality factor ($\tau/P$) lie in the range of 1.5$-$2.5. The oscillation decays much faster near the loop top than the loop leg. The parameters in 131 {\AA} are close to those in 171 {\AA}.

\begin{table}
\centering
\caption{Fitted parameters of the coronal loop oscillation.}
\label{tab-1}
\begin{tabular}{c|ccccc}
  \hline
    & $A_0$  & $P$  & $\tau$ & $\tau/P$\\
       &(Mm)  &  (s)   &  (s)  \\
     \hline
  171 {\AA} S1 & 13.6$\pm$0.3  & 479$\pm$4   & 714$\pm$26  & 1.5$\pm$0.1 \\
   \hline
   171 {\AA} S2 & 10.4$\pm$0.2 & 467$\pm$2   & 1185$\pm$49 & 2.5$\pm$0.1 \\
   \hline
   171 {\AA} S3 & 8.9$\pm$0.3  & 441$\pm$3   & 1029$\pm$55 & 2.3$\pm$0.1 \\
   \hline
   131 {\AA} S3 & 8.6$\pm$0.3  & 435$\pm$3   & 1095$\pm$66 & 2.5$\pm$0.2 \\
  \hline
\end{tabular}
\end{table}

\subsection{Magnetic field estimated from coronal seismology}
As mentioned in Sect.~\ref{s-intro}, estimation of the magnetic field strength of oscillating loops is an important application of coronal seismology.
To estimate the magnetic field of the large-scale coronal loop in Fig.~\ref{fig1}, we consider the loop as a straight cylinder with the magnetic field lines frozen.
The period of standing kink-mode oscillation depends on the loop length ($L$) and phase speed $(C_k)$ \citep{naka99}:
\begin{equation} \label{eqn-5}
  P=\frac{2L}{C_{k}}\,,
  C_{k}=\sqrt{\frac{2}{1+\rho_{e}/\rho_{i}}}C_{A},
\end{equation}
where $C_A$ is the Alfv\'{e}n speed in the loop, and $\rho_e$ and $\rho_i$ stand for the external and internal plasma densities.

\begin{figure}
\includegraphics[width=8cm,angle=-90,clip=]{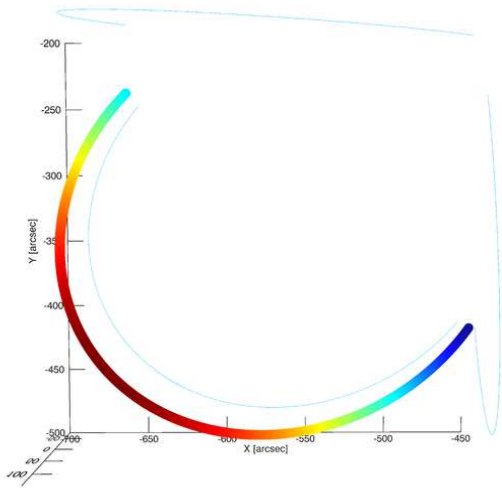}
\centering
\caption{Three-dimensional (3D) geometry of the semi-elliptical loop, with major and minor axes being 350$\arcsec$ and 264$\arcsec$.
The rotation angles of the initial loop around $x$, $y$, and $z$ axes are -60$^{\circ}$, -9$^{\circ}$, and 40$^{\circ}$, respectively.
The colors along the loop represents the heights, and the thin light blue lines represent the projections of the loop onto the three planes.}
\label{fig5}
\end{figure}

\begin{figure}
\includegraphics[width=8cm,clip=]{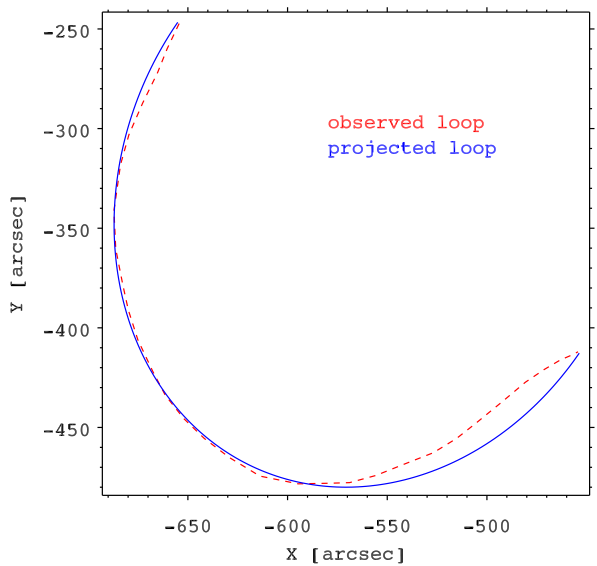}
\centering
\caption{The observed loop (red dashed line) in Fig.~\ref{fig1} and projected semi-elliptical loop in the $xy$-plane (blue solid line).}
\label{fig6}
\end{figure}

In Fig.~\ref{fig1}, the apparent distance between the footpoints of loop is $\sim$256$\arcsec$, while the real distance becomes $\sim$268$\arcsec$ after correcting the projection effect.
Since there were no stereoscopic observations from the STEREO \citep{kai05} spacecrafts, the true geometry of the loop could not be inferred by stereoscopy.
We assume a semi-elliptical shape of the loop initially in the $xz$-plane, which is determined by the major axis (2$a$) and minor axis (2$b$).
The minor axis is taken to be the length of the loop baseline in spherical coordinates, while the major axis is varied.
The loop is sequentially rotated around the $x$, $y$, and $z$ axes by angles of $\theta$, $\alpha$, and $\beta$, respectively.
The projected loop in the $xy$-plane is then translated to compare with the observed loop in Fig.~\ref{fig1}.
By varying continuously the values of 2$a$ and three rotation angles, we can find the best values of loop parameters when the average distance
between the projected loop and the observed loop is minimized using \texttt{mpfit.pro}.

Figure~\ref{fig5} shows the three-dimensional (3D) geometry of the semi-elliptical loop,
where $a=175\arcsec$, $b=132\arcsec$, $\theta=-60^{\circ}$, $\alpha\approx-9^{\circ}$, and $\beta\approx40^{\circ}$.
In Fig.~\ref{fig6}, the projected loop in the $xy$-plane is drawn with the blue solid line and the observed loop is drawn with the red dashed line.
It is clear that the projected loop fits well with the observed loop, suggesting that the semi-elliptical shape can satisfactorily represent the true geometry of the oscillating loop.
The length of the loop is calculated to be $\sim$377 Mm, which is $\sim$1.2 times longer than the value assuming a semicircular shape.
Therefore, $C_k$ is estimated to be $\sim$1630 km s$^{-1}$ by adopting an average period of oscillation ($P\approx462$ s).
$C_A$ is estimated to be $\sim$1210 km s$^{-1}$ assuming that the density ratio $\rho_e/\rho_i$ is equal to $\sim$0.1 \citep{naka99,naka01}.

\begin{figure}
\includegraphics[width=8cm,clip=]{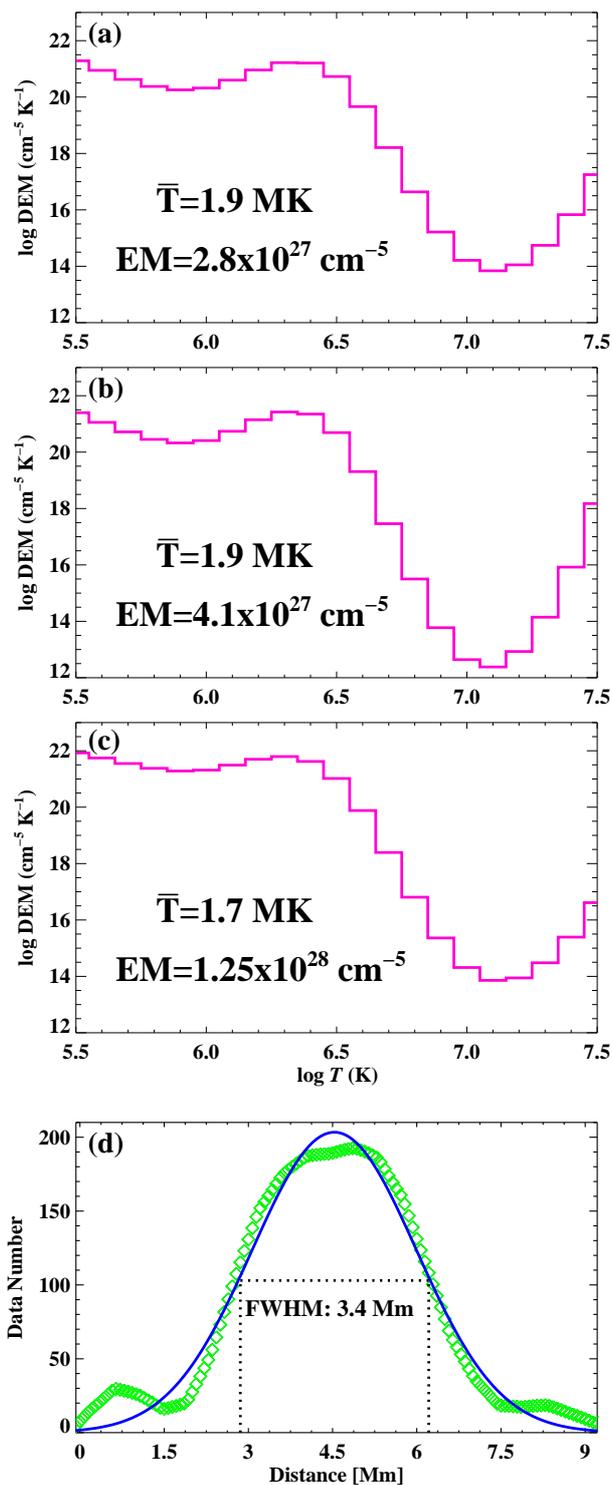}
\centering
\caption{(a-c) DEM profiles of the three tiny boxes in Fig.~\ref{fig1}(b). The average temperature and total EM are labeled.
(d) Intensity distribution of a short line across the middle box, which is fitted with a single-Gauss function.
The full width at half maximum (FWHM) representing the width or LOS depth of the loop is labeled.}
\label{fig7}
\end{figure}

The Alfv\'{e}n speed is determined by the magnetic field strength and mass density of the plasma.
Consequently, we can estimate the magnetic field strength in the loop \citep{naka01}:
\begin{equation} \label{eqn-6}
  B=\sqrt{4\pi \rho_i}C_A,
\end{equation}
where $\rho_{i}=m_{p}n_i \approx m_{p}n_e=m_{p}\sqrt{\mathrm{EM}/H}$, $m_p$ is the proton mass, and $H$ is the LOS depth of the coronal loop.
In Fig.~\ref{fig1}(b), three tiny boxes, representing the loop top, loop leg, and loop footpoint, are used to perform DEM analysis described in Sect.~\ref{s-dem}.
The inverted DEM profiles are displayed in Fig.~\ref{fig7}(a-c), with the calculated values of $\bar{T}$ and EM being labeled.
The average temperature of the oscillating loop is $\leq$2 MK, which is consistent with the fact that the oscillation is observed in 131, 171, 193, and 211 {\AA}.
The background-subtracted intensity distribution of a short line (9.1 Mm in length) across the middle box is shown in Fig.~\ref{fig7}(d).
Single-Gauss fitting of the profile is used to derive the FWHM ($\sim$3.4 Mm), which is considered as the width or LOS depth of the loop.
Hence, the number density ($n_i$) and corresponding mass density ($\rho_i$) of the loop decrease from 6$\times$10$^9$ cm$^{-3}$ and 1$\times$10$^{-14}$ g cm$^{-3}$
near the footpoint to 2.9$\times$10$^9$ cm$^{-3}$ and 4.8$\times$10$^{-15}$ g cm$^{-3}$ near the loop top.
The magnetic field strength of the loop ($B$) falls in the range of 30$-$43 G according to Equation~\ref{eqn-6}.

\subsection{Magnetic field determined by NLFFF modeling}
To validate the magnetic field strength of the oscillating loop inferred from coronal seismology,
we carry out NLFFF modeling to construct magnetic field models by using the flux rope insertion method \citep{van04}.
We have briefly introduced the method in Sect.~\ref{s-su}, and more details can be found in \citet{bob08} and \citet{su09,su11}.

\begin{figure}
\includegraphics[width=8cm,clip=]{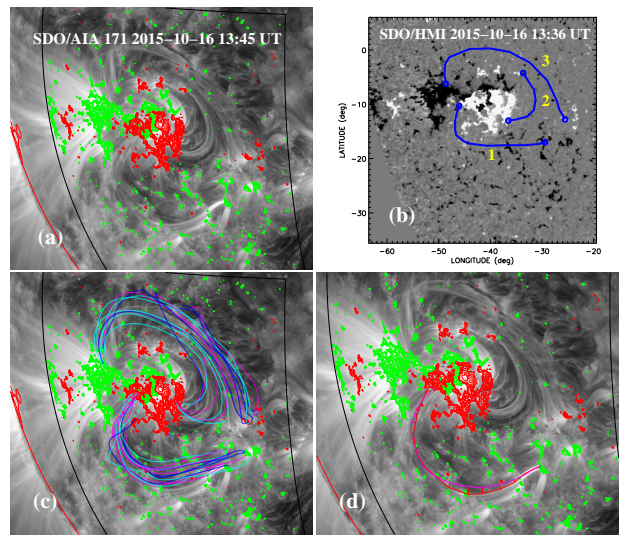}
\centering
\caption{(a) AIA 171 {\AA} image (gray scale) overlaid with positive (red contours) and negative (green contours) polarities of the photospheric magnetic field taken by HMI.
(b) Zoomed view of the longitude-latitude map of the radial component of the photospheric magnetic field by SDO/HMI in the high-resolution region.
The blue curves (1, 2, 3) with circles at the two ends refers to the three paths along which we insert the flux ropes.
(c) AIA 171 {\AA} image and selected field lines from the NLFFF model matching the observed non-potential coronal loops.
(d) AIA 171 {\AA} image and comparison of the observed coronal loop (red line) traced manually and the best-fit modeling field line (pink line).}
\label{fig8}
\end{figure}

The boundary condition for the high-resolution region is derived from the LOS magnetogram taken by HMI at 13:36 UT on 2015 October 16.
The longitude-latitude map of the radial component of the magnetic field in the high-resolution region is presented in Fig.~\ref{fig8}(b).
Three flux ropes with the same poloidal flux (0 Mx cm$^{-1}$) are inserted.
However, the axial fluxes are different, namely, 8$\times$10$^{20}$ Mx, 1$\times$10$^{20}$ Mx and 4$\times$10$^{20}$ Mx, respectively.
In Fig.~\ref{fig8}(c), selected model field lines matching the observed non-potential coronal loops are overlaid on the AIA 171 {\AA} image.
In Fig.~\ref{fig8}(d), the observed loop is traced manually and marked with the red line.
In order to find the field line that best fit the observed loop, we first measure the distance between a point on the observed loop and the closest point on the projected field line in the image plane.
These distances for various points along the observed loop are then averaged, which is defined as the ``average deviation" \citep{su09}.
The manually selected 3D field line that can minimize the deviation is considered as the line that best fits the observed loop (pink line), which is overlaid on the AIA 171 {\AA} image.
The length of the pink line ($\sim$354 Mm) accounts for $\sim$94\% of the loop length assuming a semi-elliptical shape.
We obtained the magnetic field strength at several locations along the model field line,
which lies in the range of 21$-$23 G and is in the same order of magnitude as the result of coronal seismology.

\section{Summary and Discussion} \label{s-sum}
In this work, we report our multiwavelength observations of the transverse oscillation of a large-scale coronal loop induced by a blowout jet related to a C4.2 CRF in AR 12434 on 2015 October 16.
The oscillation is most pronounced in AIA 171 and 131 {\AA}.
The oscillation is almost in phase along the loop with a peak initial amplitude of $\sim$13.6 Mm, meaning that the oscillation belong to the fast standing kink mode.
The oscillation lasts for $\sim$3.5 cycles with an average period of $\sim$462 s and average damping time of $\sim$976 s. The values of $\tau/P$ lie in the range of 1.5$-$2.5.
Based on coronal seismology, the Alfv\'{e}n speed in the oscillating loop is estimated to be $\sim$1210 km s$^{-1}$.
Two independent approaches are applied to calculate the magnetic field strength of the loop,
resulting in 30$-$43 G using the coronal seismology and 21$-$23 G using the NLFFF modeling.
The results of two methods are in the same order of magnitude, which confirms the reliability of coronal seismology in diagnosing coronal magnetic field.

As mentioned in Sect.~\ref{s-intro}, transverse loop oscillations triggered by CRFs have been observed.
\citet{zqm15} analyzed an M6.7 flare as a result of partial filament eruption on 2011 September 8. Kink oscillation was induced in an adjacent coronal loop within the same AR (see their Fig. 7).
The oscillation with a small amplitude ($\sim$1.6 Mm) lasted for $\sim$2 cycles without significant damping. The estimated parameters are listed and compared with this study in Table~\ref{tab-2}.
It is found that the values of $C_k$ and $C_A$ are close to each other for the two events. Both the loop length and period in this study are $\ge$2 times larger than those for the event in 2011.
However, the density and magnetic field of the loop in 2011 are much larger than those in our study.
\citet{zqm20} investigated the decayless and decaying kink oscillations of an EUV loop on 2014 March 5.
For comparison, the parameters, including the shortest loop length and period among the three events, are also listed in the last row of Table~\ref{tab-2}.

\begin{table*}
\centering
\caption{Parameters of the oscillating coronal loops observed by SDO/AIA on 2011 September 8, 2015 October 16, and 2014 March 5.}
\label{tab-2}
\begin{tabular}{ccccccccc}
\hline
 Date  & AR & $P$ & $L$ & $C_k$  & $C_A$ & $n_e$ & $B$ & Ref.\\
    & &  (s)  &  (Mm)  &  (km s$^{-1}$) & (km s$^{-1}$) & (10$^9$ cm$^{-3}$) & (G) &  \\
\hline
2011/09/08 & 11283 & 225 & 167 & 1482 & 1100 & 25 & 79 & \citet{zqm15} \\
2015/10/16 & 12434 & $\sim$462 & $\sim$377 & $\sim$1630 & $\sim$1210 & 3$-$6 & 30$-$43 & this study \\
2014/03/05 & 11991 & $\sim$117 & 130 & 2200 & 1555 & 7$-$10 & 65$-$78 & \citet{zqm20} \\
\hline
\end{tabular}
\end{table*}

\citet{as11} compared the magnetic field of an oscillating loop determined by coronal seismology with the result of magnetic extrapolation based on the potential field source surface (PFSS)
modeling using the photospheric magnetogram. It is found that the average extrapolated magnetic field strength exceeded the seismologically determined value ($\sim$4 G) by a factor of two.
After improving the method of estimation of physical parameters by taking the effect of density and magnetic stratification into account, the extrapolated magnetic field are optimized \citep{ver13a}.
\citet{guo15} reported the kink oscillation of a coronal loop with a total length of $\sim$204 Mm,
which was excited by the global fast magneto-acoustic wave as a result of flux rope eruption and the associated eruptive flare on 2013 April 11.
Based on coronal seismology, they derived the spatial distribution of magnetic field ($\sim$8 G) along the loop, which matches with that derived by a potential field model.
\citet{Long2017} estimated the magnetic field of a trans-equatorial loop system using two independent techniques.
It is found that the magnetic field strength ($\sim$5.5 G) estimated by two approaches are roughly equal.

\begin{acknowledgements}
The authors are grateful to the referee for valuable suggestions. The authors also appreciate Drs. Z. J. Ning, D. Li, and F. Chen for valuable discussions.
SDO is a mission of NASA\rq{}s Living With a Star Program. AIA and HMI data are courtesy of the NASA/SDO science teams.
Q.M.Z. is supported by the Science and Technology Development Fund of Macau (275/2017/A), CAS Key Laboratory of Solar Activity, National Astronomical Observatories (KLSA202006),
Youth Innovation Promotion Association CAS, and the International Cooperation and Interchange Program (11961131002).
This work is funded by NSFC grants (No. 11790302, 11790300, 11773079, 41761134088, 11473071, 12073081),
and the Strategic Priority Research Program on Space Science, CAS (XDA15052200, XDA15320301).
\end{acknowledgements}

\end{document}